\documentclass[prl,twocolumn,showpacs,groupedaddress]{revtex4}

\usepackage{dcolumn}
\usepackage{graphicx}
\usepackage{amsmath}
\usepackage{bm}
\usepackage{amssymb}
\usepackage{wasysym}

\usepackage[usenames]{color}

\begin{document}

\title{Nonclassical Radiation from Thermal Cavities in the Ultrastrong Coupling Regime}
\author{ A. Ridolfo$^1$, S. Savasta$^2$, and M. J. Hartmann$^1$}
\affiliation{$^1$Physik Department, Technische Universit\"{a}t M\"{u}nchen, 85748 Garching, Germany\\$^2$Dipartimento di Fisica  e Scienze della Terra, Universit\`{a} di Messina, I-98166 Messina, Italy} 
\date{\today}
%\maketitle
%%%%%%%%%%%%%%%%%%%%%%%%%%%%%%%%%%%%%%%%%%%%%%%%%%%%%%%%%
\begin{abstract}
{Thermal or chaotic light sources emit radiation characterized by a slightly enhanced probability of emitting photons in bunches, described by a zero-delay second-order correlation function $g^{(2)}(0) = 2$. Here we explore photon-coincidence counting statistics of thermal cavities in the ultrastrong coupling regime, where the atom-cavity coupling rate becomes comparable to the cavity resonance frequency. We find that, depending on the system temperature and coupling rate, thermal photons escaping the cavity can display very different statistical behaviors, characterised by second-order correlation functions approaching zero or greatly exceeding two.}
\end{abstract}

\pacs{42.50.Pq, 42.50.Ar, 85.25.-j, 03.67.Lx}

\maketitle

Thermal radiation has a special place in modern physics. In the search for a solution to the discrepancies between the observed energy spectrum of thermal radiation and the predictions of classical theory, Planck was led to introduce the revolutionary concept of quanta \cite{Planck}. Thermal or chaotic light sources emit radiation that is characterized by an enhanced probability of emitting photons in bunches \cite{HBT}. In the course of the successful attempt to explain this effect, Glauber established the basis of quantum optics \cite{Glauber, NobelGlauber}.
Even more recently the study of thermal emission has continued to provide amazing results.
A thermal light-emitting source is often presented as a typical example of an incoherent source. However, it has been shown  that the field emitted by a thermal source made of a polar material is enhanced by more than four orders of magnitude and displays first-order coherence in the near-field zone \cite{GreffetPRL}. Moreover, by introducing a periodic microstructure into such a polar material, a thermal infrared source can be fabricated that displays first-order coherence over large distances \cite{GreffetNature}. While first-order coherence and spectral properties of thermal sources can be manipulated and tailored, their second-order coherence is known to be completely absent \cite{NobelGlauber} resulting in the small bunching described by $g^{(2)}(0) = 2$.

Here we investigate  photon-coincidence counting statistics of thermal cavities in the ultrastrong coupling regime, where the strength of the interaction $g$ between an emitter and the cavity photons becomes comparable to the transition frequency of the emitter $\omega_{\rm x}$ or the frequency of the cavity mode $\omega_0$. Ultrastrong light-matter interactions have recently been achieved both in semiconductor and superconducting systems where exceptionally high field amplitudes for the photons and emitters with very large dipole moments can be realized \cite{guenter09,Niemczyk,Todorov,Schwartz,Hoffman, Scalari}. These structures are attracting increasing interest due to the possibility of manipulating the physical properties of the cavity quantum electrodynamic ground state. 

The photon statistics of chaotic sources (like thermal cavities) and lasers can usually be explained classically. In contrast, strongly nonlinear photonic systems can emit individual photons well separated in time from each other when excited coherently or operating very far from thermal equilibrium. 
For the systems considered so far, such a scenario is however known to not persist when the coupled system is driven by thermal noise induced by reservoirs at finite temperature. Indeed, the standard quantum optics master equation (ME), generally used to study the dynamics of cavity QED systems \cite{Carmichael}, predicts for such systems $g^{(2)}(0)= 2$ independently of temperature and coupling strength. The interaction between atoms and cavity photons is most often neglected when considering the coupling of this system to an environment. Recently it has however been shown that this simplification, which leads to the standard quantum optics ME, can generate unphysical effects in the ultrastrong coupling regime \cite{Blais}. Another key issue is the failure of standard quantum optical normal order correlation functions to describe photodetection experiments for such systems \cite{DeLiberato2,RidolfoPrl2012}. The theoretical treatment of this regime thus requires a description that goes beyond the standard techniques of quantum optics.  

Exploiting generalized correlation functions as introduced in \cite{RidolfoPrl2012} and a ME that fully takes into account the qubit-resonator coupling \cite{Davies1974,Blais}, we investigate the photon-coincidence counting statistics of thermal sources for arbitrary light-matter coupling.
For this purpose we consider a single mode resonator coupled to a two-level quantum emitter where each subsystem is coupled to independent thermal baths of harmonic oscillators at a common temperature $T$. We concentrate on the zero-detuning $\omega_0 =\omega_{\rm x}$ and low-temperature cases, where more striking deviations from the standard results appear. 
We moreover have corroborated the generality of our findings by confirming that similar results are found for one cavity mode coupled to multiple emitters or one emitter coupled to multiple cavity modes, see supplementary material \cite{supplement}.

A particularly well suited technology for such an experiment are superconducting circuits \cite{wallraff04,Norireview} which have recently emerged as an excellent platform for microwave on-chip quantum-optics experiments and where second-order correlation function measurements for quantum \cite{menzel10,Bozyigit,Lang} and low temperature thermal fields \cite{Mariantoni10} have been performed using quadrature amplitude detectors. 

\paragraph{Model -}
The cavity QED system we explore consists of a single-mode resonator that interacts with a two level system (TLS). This system can be described by the Rabi Hamiltonian (assuming $\hbar = 1$),
\begin{equation} \label{eq:model}
       H = \omega_{\rm 0} a^{\dagger} a + \omega_{\rm x} \sigma^{+} \sigma^{-} + g ( a + a^{\dagger}) (\sigma^{-} + \sigma^{+})
\end{equation} 
where $\omega_{\rm 0}$ and $\omega_{\rm x}$ are the cavity and TLS bare frequencies, $g$ is the coupling strength, while $a$ $(a^{\dagger})$ denotes the annihilation (creation) operator for the cavity and $\sigma^{-}$ $(\sigma^{+})$ is the lowering (raising) operator for the TLS. Since we are interested in probing photon-coincidence statistics for arbitrary light-matter interactions where the contribution of counter rotating terms cannot be neglected, we do not make use of the rotating wave approximation (RWA) \cite{asquarenote}. Recently, it was shown that in the limit of very large coupling, the ground state and the first excited state of this system can become quasidegenerate \cite{Nataf,noripra2010,Emary}.

In order to study thermal emission as well as the statistics of thermal photons we calculate the normal-order correlation functions of the output field. Standard normal order correlation functions were recently shown to not correctly describe the emission properties and photon statistics of systems in the ultrastrong coupling regime \cite{RidolfoPrl2012} as they would, for example, predict an unphysical stream of output photons even for a zero-temperature system, $\langle a^\dag a \rangle_{{}_{T=0}} = \text{Tr}[a^\dag a \, \rho_{{}_{T=0}}] \neq 0$. Following \cite{RidolfoPrl2012}, we here employ correlation functions for the output fields that are valid for an arbitrary coupling strength by expressing the cavity field $X = -i X_0 (a - a^\dag)$
($X_0$ is the  rms zero-point field-amplitude) in the atom-cavity dressed basis. In particular, the set-up we have in mind is equivalent to the emission of a thermalized black-box, whose output is coupled to the vacuum of a one-dimensional waveguide. In this case, the output and input operators, $a_{\text{out}}(t)$ and $a^{\text{vac}}_{\text{in}}(t)$, obey the relation,  
%
%\begin{equation} \label{eq:input-output}
$a_{\text{out}}(t) = a^{\text{vac}}_{\text{in}}(t) - i \sqrt{\gamma_{a}} \dot{X}^{+}$ \cite{RidolfoPrl2012}.
%\end{equation}
%
Here $X^+$ ($X^- = (X^+)^\dag$) is the positive (negative) frequency component of the cavity field $X$, which
can be derived by expanding the respective operators in the dressed state basis, namely the eigenstates
$| j \rangle$ of $H$ as in Eq. (\ref{eq:model}) ordered according to increasing eigenenergies $\omega_j$. 
Specifically, the time derivative  of $X^{+}$ can be expressed as $\dot{X}^{+} = -i \sum_{j,k>j} \Delta_{kj}X_{jk} | j \rangle \langle k |$ where $X_{jk} = \langle j | X | k \rangle$ and $\Delta_{kj} = \omega_k - \omega_j$.
According to these input-output relations
and for input fields in vacuum, the normalized second order correlation function for the output field reads,
\begin{equation} \label{eq:g2ofX}
g^{(2)}(\tau) = \lim_{t \to \infty} \frac{\langle \dot{X}^{-}(t) \dot{X}^{-}(t+\tau) \dot{X}^{+}(t+\tau) \dot{X}^{+}(t) \rangle}{\langle \dot{X}^{-}(t) \dot{X}^{+}(t) \rangle^{2}}\, .
\end{equation}

\paragraph{Results -}
The thermal-equilibrium  zero-delay correlation function $g^{(2)}(0)$ can be directly calculated from the thermal equilibrium density-operator \cite{note}. For a system in thermal equilibrium, statistical properties are related to the density matrix of the canonical ensemble, $\rho_{T}$, that is the most general way to describe such thermalized interacting systems. In the basis where $H$ is diagonal, $\rho_{T}$ reads,   
\begin{equation} \label{eq:canonical}
\rho_{T} = \frac{e^{-(\epsilon_{j}/k_{B}T)}}{\cal{Z}} \delta_{i j},
\end{equation}
where $\epsilon_{j}$ is the \emph{j-}th eigenvalue of $H$, $\delta_{i j}$ the Kronecker delta, and ${\cal Z} = \sum_{j} \exp(-\epsilon_{j}/k_{B}T)$ the partition function.

Figure 1 shows the thermal-equilibrium  zero-delay correlation function $g^{(2)}(0)$ as a function of the effective coupling $g/ \omega_0$ and temperature for zero detuning ($\omega_{\rm 0} = \omega_{\rm x}$). 
%
%========================================================================
\begin{figure}
\centering
\includegraphics[width=\columnwidth]{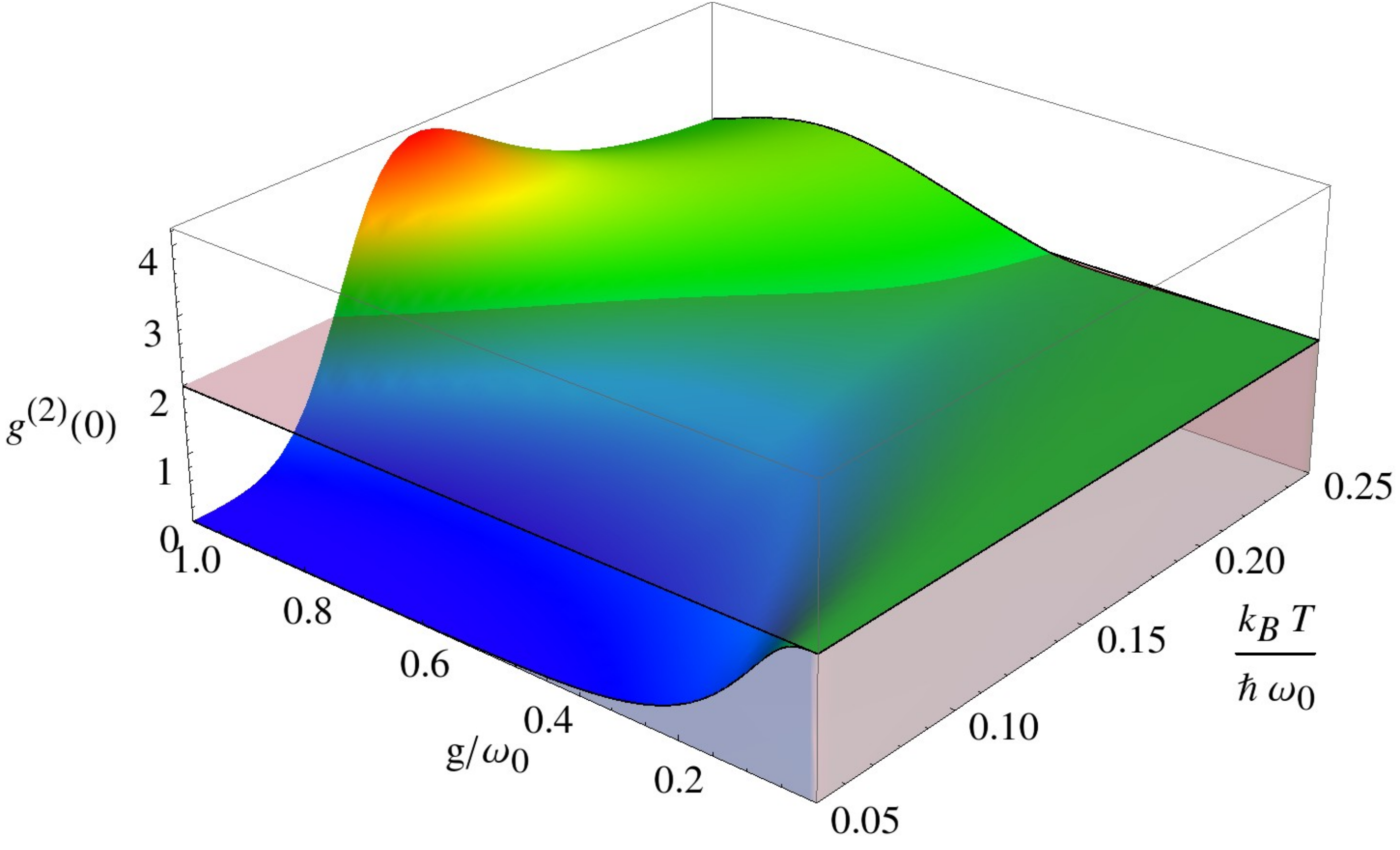}
\caption{(color online) $g^{(2)}(0)$ plotted as a function of the temperature and coupling strength. The results are obtained on resonance ($\omega_{\rm 0} = \omega_{\rm x}$) and for a steady-state with cavity and TLS in thermal equilibrium, i.e. $T_{a} = T_{x}$. Notably, in thermal equilibrium, statistical properties are independent of the damping rates. For comparison we show, the plane of $g^{(2)}(0) = 2$ which indicates the value that would result from a conventional ME, where a RWA is performed.}
\label{fig:g20}
\end{figure}
%========================================================================
The calculated $g^{(2)}(0)$ exhibits striking differences from the standard value of $g^{(2)} = 2$. 
Of particular interest is the region with large effective coupling $g/\omega_0 > 0.4$ and low temperature $k_B T /\omega_0 < 0.1$ where $g^{(2)}(0) \ll 1$. This highly nonclassical behavior of thermal photons opens prospects towards the realization of thermal sources of single photons in circuit QED. This anomalous behavior originates from the tendency of the interacting quantum system  towards vacuum degeneracy  for large couplings. Specifically for increasing coupling the energy of the first excited state converges towards that of the ground state, while the other energy levels remain well separated from that doublet (see Fig. 2b). Hence, at sufficiently low temperature, only the first excited state is significantly populated by thermal noise ($(\omega_{i} - \omega_0)/ kT$ is non-negligible only for $i =1$). At the onset of vacuum degeneracy, the ground and first excited state
are quantum superpositions with multi-photon components \cite{Nataf} and naively one might expect to observe bunching effects. However such photons are mostly virtual and the use of generalized normal order correlation functions shows that the transition $| 1 \rangle \to | 0 \rangle$ can only emit one physical photon at a time.
%
%========================================================================
\begin{figure}
\centering
\includegraphics[height=53mm]{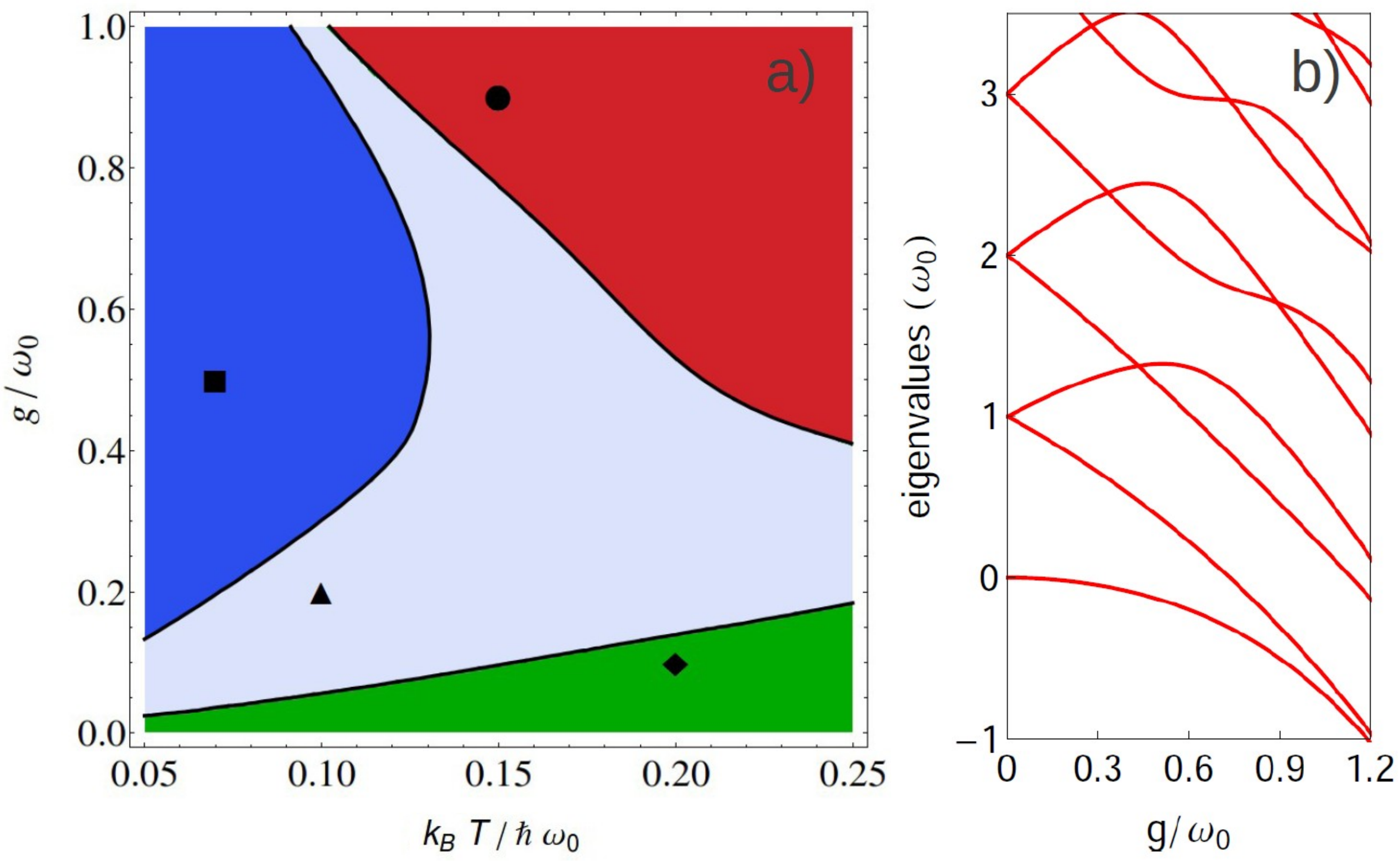}
\caption{(color online) {\bf a}) Contourplot of $g^{(2)}(0)$ calculated with the same parameters as Fig. \ref{fig:g20}. Here one can distinguish four regions: i) green, with $1.999< g^{(2)}(0) < 2$ corresponding to the standard thermal result, ii) gray, for $1< g^{(2)}(0) < 1.999$, iii) blue, with the sub-Poissonian values $g^{(2)}(0) < 1$ and iv) red, with a $g^{(2)}(0) > 2$. The markers identify data points that we investigate further to characterise the different behaviors of $g^{(2)}(\tau)$. Their ($g/\omega_{0} , k_{B}T/\hbar \omega_{0}$) values are: $\blacklozenge$ $(0.1,0.2)$, $\blacktriangle$ $(0.2,0.1)$, $\blacksquare$  $(0.5,0.07)$, $\CIRCLE$ $(0.9,0.15)$. {\bf b}) Energy eingenvalues of $H$ as in Eq. (\ref{eq:model}) as a function $g/\omega_{0}$.}
\label{fig:contourplot}
\end{figure}
%========================================================================

For large effective couplings and higher temperatures $g^{(2)}(0)$ becomes larger than the standard value, $g^{(2)}(0) = 2$. 
Also this behaviour can be understood from the spectrum of the Hamiltonian $H$ in Eq. (\ref{eq:model}). As $H$ conserves the parity of the number of TLS- and cavity-excitations, its eigenstates either contain an odd or an even number of such excitations but not both. The single excitation decays associated with $\dot{X}^{+}$ can thus only occur between eigenstate with different parities. As the energy levels that converge to $n \omega_{0}$ for $g\to 0$, c.f.  Fig. \ref{fig:contourplot}b, contain $n$ excitations, the decay process $|2\rangle \to |1\rangle$ can not occur for $g/\omega_0\ll 1$. At $g/\omega_0 \simeq 0.45$ however a crossing between the second and third excited level occurs and beyond this point the cascaded decay $|2\rangle \to |1\rangle \to |0\rangle$ is possible \cite{expl}. For temperatures that are sufficiently high to appreciably populate these levels, such cascaded decays can lead to pronounced bunching with $g^{(2)}(0) > 2$, c.f. Fig. \ref{fig:g20}. Since the energies of excited states decrease as $g/\omega_0$ grows, their population increases for a given temperature.

An overview of the behaviors of $g^{(2)}(0)$ is shown in the contour plot of Fig. \ref{fig:contourplot}, where four different regions are displayed:
i) a region for small $g/\omega_{0}$ (green) where the standard thermal result $g^{(2)} \approx 2$ is recovered (here $1.999< g^{(2)}(0) < 2$) and which broadens for increasing temperatures; ii) a sub-Poissonian or nonclassical region with $g^{(2)} < 1$ (blue) in the ultrastrong coupling regime and for sufficiently low temperatures; iii) a region with an intermediate regime (gray) with $1 < g^{(2)}(0) < 2$; and iv) a super-Poissonian region (red) beyond the standard value of $g^{(2)} = 2$ for very large coupling and higher temperatures. We note here that a calculation of $g^{(2)}(0)$ within the RWA and using the standard quantum optics ME \cite{Carmichael} would always yield the value $g^{(2)}(0) = 2$, regardless of the coupling strength and the temperature.

While zero-delay correlation function can be directly inferred from the thermal density operator, a description of the time dependent dynamics of the open quantum system is required to calculate the time-delayed second order correlation function $g^{(2)}(\tau)$. A viable description of system-bath interactions typically requires an expansion in the system bath coupling. A suitable way to perform this perturbative expansion consists in writing the Hamiltonian in the basis of its eigenstates $|j\rangle$ \cite{Davies1974}. In this way we obtain the following ME \cite{breuer,Blais},
\begin{equation}\label{eq:master-eq}
    \dot\rho(t) = i [\rho(t), H] + \mathcal{L}_{a}\rho(t) + \mathcal{L}_{x}\rho(t),
\end{equation}
where $\mathcal{L}_{a}$ and $\mathcal{L}_{x}$ are Liouvillian superoperators describing the losses and the thermal feeding of the system. They read,
$ \mathcal{L}_{c}\rho(t) = \sum_{j,k>j}\Gamma^{j k}_{c}\bar{n}(\Delta_{k j},T)\mathcal{D}[|k \rangle \langle j|]\rho(t) + \sum_{j,k>j}\Gamma^{j k}_{c}(1 + \bar{n}(\Delta_{k j},T))\mathcal{D}[|j \rangle \langle k|]\rho(t)$ for $c = a, \sigma^{-}$ with $\mathcal{D}[\mathcal{O}]\rho = \frac{1}{2} (2 \mathcal{O}\rho\mathcal{O}^{\dagger}-\rho \mathcal{O}^{\dagger} \mathcal{O} - \mathcal{O}^{\dagger} \mathcal{O}\rho)$. 
The relaxation coefficients $\Gamma^{j k}_{c} = 2\pi d_{c}(\Delta_{k j}) \alpha^{2}_{c}(\Delta_{k j})| C^{c}_{j k}|^2$ depend on the spectral density of the baths, $d_{c}(\Delta_{k j})$, and the system-bath coupling strength, $\alpha_{c}(\Delta_{k j})$, at the respective transition frequency $\Delta_{k j} = \omega_{k} - \omega_{j}$ as well as on the transition coefficients $C_{j k} = -i \langle j |(c - c^{\dagger})| k \rangle$
($c = a, \sigma^{-}$). $\bar{n}(\Delta_{k j},T)$ is the thermal population at frequency $\Delta_{k j}$ and temperature $T$. 
In all examples reported in this work, we consider a cavity that couples to the momentum quadratures of fields in one-dimensional output waveguides, assuming that the spectral density $d_{c}(\Delta_{k j})$ is constant and $\alpha_{c}^{2}(\Delta_{k j}) \propto \Delta_{k j}$. Hence the relaxation coefficients reduce to $ \Gamma^{j k}_{c} = \gamma_{c} \,(\Delta_{k j}/\omega_{0}) \, |C^{c}_{j k}|^2$, where $\gamma_{c}$ are the standard damping rates.
These assumptions correspond to typical experimental settings e.g. in circuit-QED. In the ME (\ref{eq:master-eq}) we neglect contributions of dephasing noise and Lamb shifts as they do not affect significantly our findings.
A first result worth mentioning is that the steady state solution of Eq.\ (\ref{eq:master-eq}) reproduces the thermal state of Eq. (\ref{eq:canonical}) that is independent of the damping rates. This confirms that the present approach is able to correctly describe the system's thermalization and thus corroborates the accuracy of Eq.\ (\ref{eq:master-eq}). We solve Eq.  (\ref{eq:master-eq}) using $\rho_{T}$ as initial condition to calculate $g^{(2)}(\tau)$ via the quantum regression theorem \cite{GardinerZoller}.
%
%========================================================================
\begin{figure}
\centering
\includegraphics[width=0.9\columnwidth]{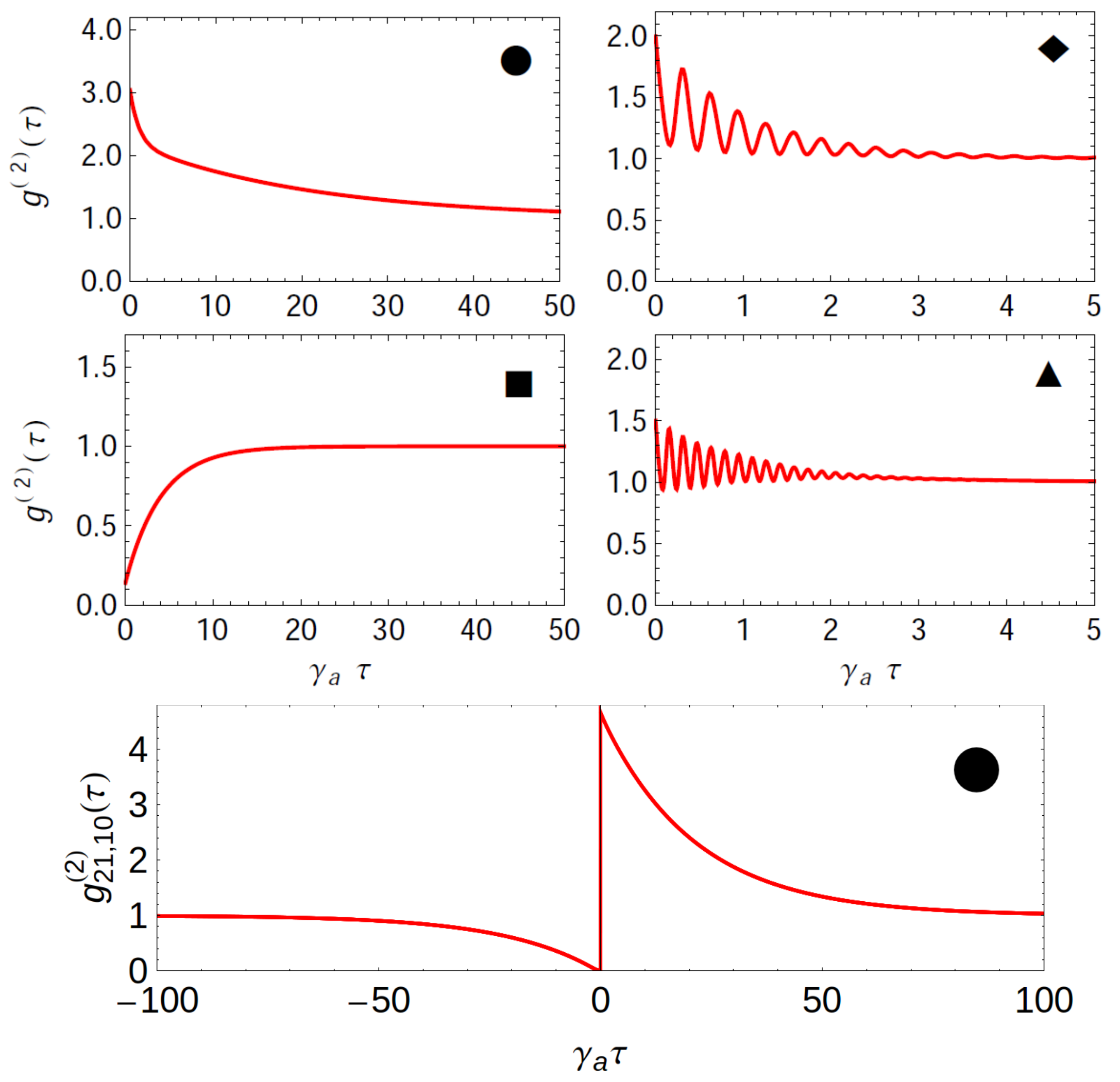}
\caption{(color online) $g^{(2)}(\tau)$ (upper panel) and $g^{(2)}_{\rm 21,10}(\tau)$ (lower panel) calculated for couplings and temperatures corresponding to the markers in Fig. \ref{fig:contourplot}. Here the damping rates are $\gamma_{a} = \gamma_{x} = 0.01 \omega_{0}$.} 
\label{fig:g2tau}
\end{figure}
%========================================================================
%

\begin{figure}
\centering
\includegraphics[width=0.9\columnwidth]{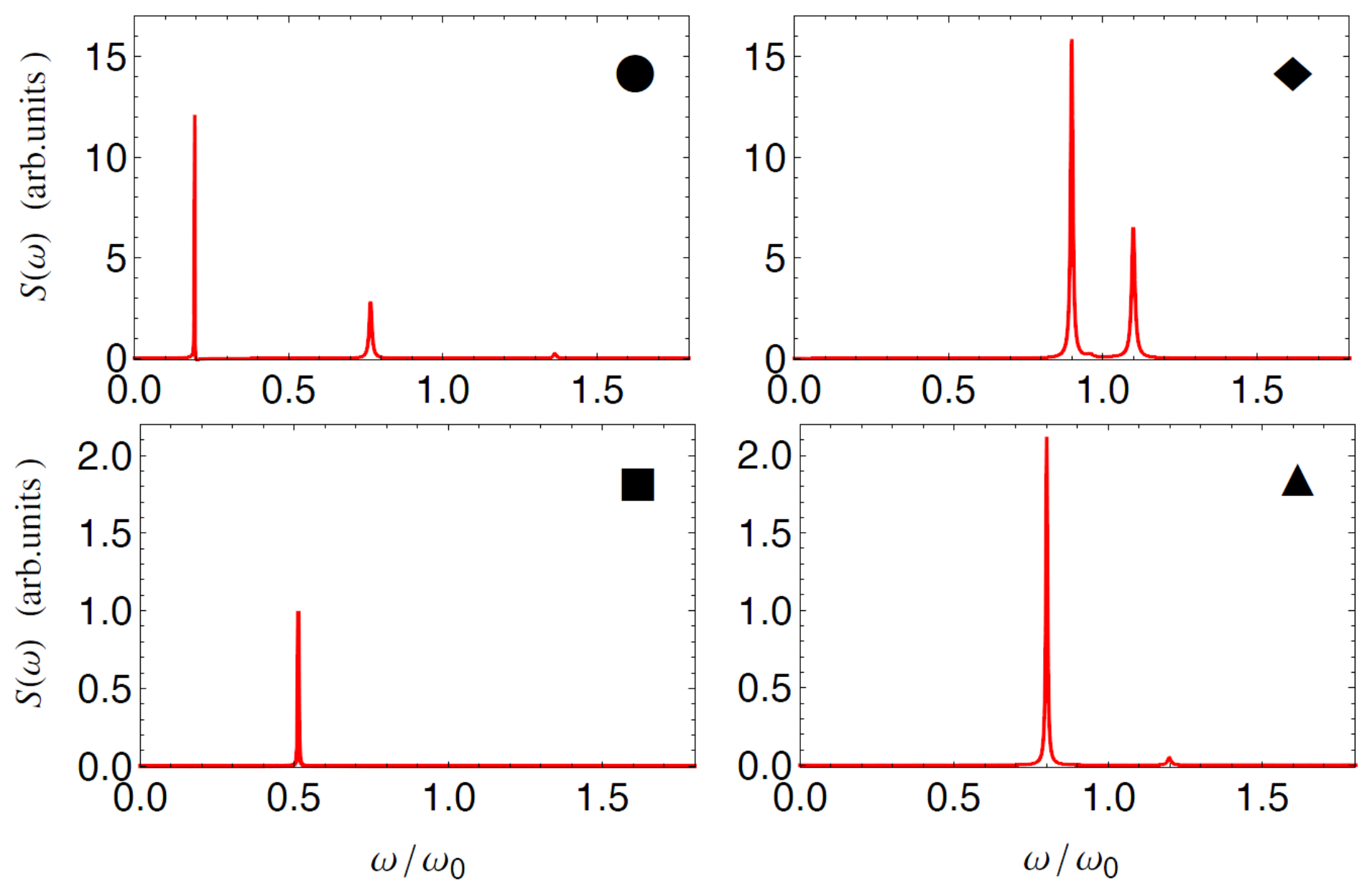}
\caption{(color online) Emission spectra calculated for couplings and temperatures corresponding to the markers in Fig. \ref{fig:contourplot} and for $\gamma_{a} = \gamma_{x} = 0.01 \omega_{0}$. Spectra heights are normalized by the maximum value of the lowest temperature spectrum (square marker).}
\label{fig:spectra}
\end{figure}

Figure \ref{fig:g2tau} displays $g^{(2)}(\tau)$, for the temperatures and couplings corresponding to the reported markers in Fig. \ref{fig:contourplot} and $\gamma_{a} = \gamma_{x} = 0.01 \omega_{0}$. 
Notably, oscillations appear in the gray and green regions. These arise from interferences of the possible decay channels. In fact, for these values of the coupling strength, the separation between the energy levels is not high enough to suppress thermal occupation of higher excited states. Hence decays from upper energy manifolds into lower excited states are possible. In this way, e.g. the third excited energy level can decay into the second or into the first, and interferences of these possible decays result in the observed oscillating behavior. This explanation is corroborated by the fact that the frequency of the oscillations corresponds exactly to $\Delta_{12}$, i.e. the energy difference between the first and the second excited state of $H$. Yet, in the blue region, $g^{(2)}(\tau)$ is almost zero and behaves as for a usual TLS, moreover the oscillations in $\tau$ are suppressed due to the low thermal noise and great separation between the energy eigenvalues. Instead, in the red region, $g^{(2)}(\tau)$ shows a slow decay for large $\tau$ which is due to the fact that the 
lifetime of an excitation in the state $| 1 \rangle$ is a lot longer than of an excitation in the state $| 2 \rangle$. This difference in lifetimes is obvious from widths of the corresponding lines in the spectrum, see Fig. \ref{fig:spectra}. Hence after the emission of a photon from the decay $| 2 \rangle \to | 1 \rangle$, the probability of detecting a second photon that originates from the decay $| 1 \rangle \to | 0 \rangle$ is substantial for a long delay range $\tau$. For this reason, $g^{(2)}(\tau)$ in the red region remains bunched for such a long time. This behavior is a consequence of the spectral density of the bath that, being a linear function of the frequencies, tends to narrow the spectral linewidth of the lower resonances.

Deeper insights into the cascade processes involving transitions with different frequencies can be obtained from the frequency filtered second-order cross-correlation functions \cite{Moreau,delValle2012}. Here we calculate them for the circle-marker case considering the frequencies $\Delta_{2,1}$ and $\Delta_{1,0}$,
\begin{equation*} %\label{eq:filteredg2}
    g^{(2)}_{\rm 12,10}(\tau) = \lim_{t \to \infty} \frac{\langle {\cal T} \dot{X}^-_{21}(t)\dot{X}^-_{10}(t + \tau)\dot{X}^+_{01}(t + \tau)\dot{X}^+_{12}(t)  \rangle}{\langle \dot{X}^-_{21}(t) \dot{X}^+_{12}(t)  \rangle \langle \dot{X}^-_{10}(t) \dot{X}^+_{01}(t)  \rangle}\, ,
\end{equation*}
where ${\cal T}$ is the time-ordering operator that positions the operators at earlier times on the outer places in the expectation value and $\dot{X}_{jk}^{+} = i  \Delta_{jk} X_{jk} | j \rangle \langle k |$ with $k>j$. 
For $\tau > 0$, $g^{(2)}_{\rm 21,10}(\tau)$ shows pronounced bunching due to the cascaded decay process $|2 \rangle \to |1\rangle \to |0\rangle$. For $\tau < 0$ on the other hand, $g^{(2)}_{\rm 21,10}(\tau)$ is anti-bunched as after detecting a photon with frequency $\Delta_{1,0}$, the system needs to be re-excited before a photon at frequency $\Delta_{2,1}$ can be emitted and detected \cite{step}.

Thermal emission is also characterized by its power spectrum, i.e. the Fourier transform of the two time correlation $S(\omega) \propto \lim_{t \to \infty} 2 \Re \int_{0}^{\infty} \langle \dot{X}^{-}(t)\dot{X}^{+}(t+\tau)\rangle e^{i \omega \tau} d\tau$.
%
%========================================================================
%========================================================================
%
We exploit again the quantum regression theorem to calculate the relevant two-time correlation function. Fig. \ref{fig:spectra} shows $S(\omega)$ for the same parameters as used in Fig. \ref{fig:g2tau} and the corresponding values of temperature and couplings for the different markers as introduced in the caption of Fig. \ref{fig:contourplot}. As expected the heights of the spectra increase for increasing temperature. Moreover the resonances have i) different linewidths, as one can see from the definition of the damping rates and ii) different heights. The latter is mainly a consequence of the thermal feeding. For a fixed temperature, the thermal occupation of a spectral resonance (i.e. its height) is determined by $\bar{n}(\omega,T)$ that increases as the frequency $\omega$ of the resonance decreases. The presence of two different decay times in the example with the dot marker becomes apparent via the different linewidths of the resonances contributing to the signal, see Fig. \ref{fig:spectra}. 
Finally, the effects described here could also be measured by populating the system  with a  quasithermal field distribution realized by mixing a fixed frequency microwave tone with  noise sources of different bandwidths \cite{Lang}.
\paragraph{Acknowledgements -}
SS acknowledges useful discussions with Omar Di Stefano and Roberto Stassi.
This work is part of the Emmy Noether project HA 5593/1-1 and the CRC 631, both funded by the German Research Foundation, DFG.

%%%%%%%%%%%%%%%%%%%%%%% References %%%%%%%%%%%%%%%%%%%%%%%%%

\widetext
\section{Supplementary Material}

Here we report calculations of $g^{(2)}(0)$ for many TLSs coupled to a single cavity mode, and for a single TLS coupled to two cavity modes. The latter case is of particular interest because of the unavoidable coupling of the TLS to higher modes that is often present in a real experimental setup. In each graph we plot $g^{(2)}(0)$ as function of the temperature and of coupling strength, imposing $\omega_{\rm 0} = \omega_{\rm x}$ and $T_{a} = T_{x}$. We show moreover in all the graphs the plane $g^{(2)}(0) = 2$ to highlight the differences to the standard ME calculation which applies a RWA. The generalization of our model to many TLSs is straightforward, and described by the Hamiltonian
$H = \omega_{\rm 0} a^{\dagger} a + \sum_{j} \omega_{\rm x}^{(j)} \sigma^{+}_{\rm j} \sigma^{-}_{\rm j} + (a + a^{\dagger})\sum_{j}g^{(j)}(\sigma^{-}_{\rm j} + \sigma^{+}_{\rm j})$.
 %========================================================================
\begin{figure}
\centering
\includegraphics[height=53mm]{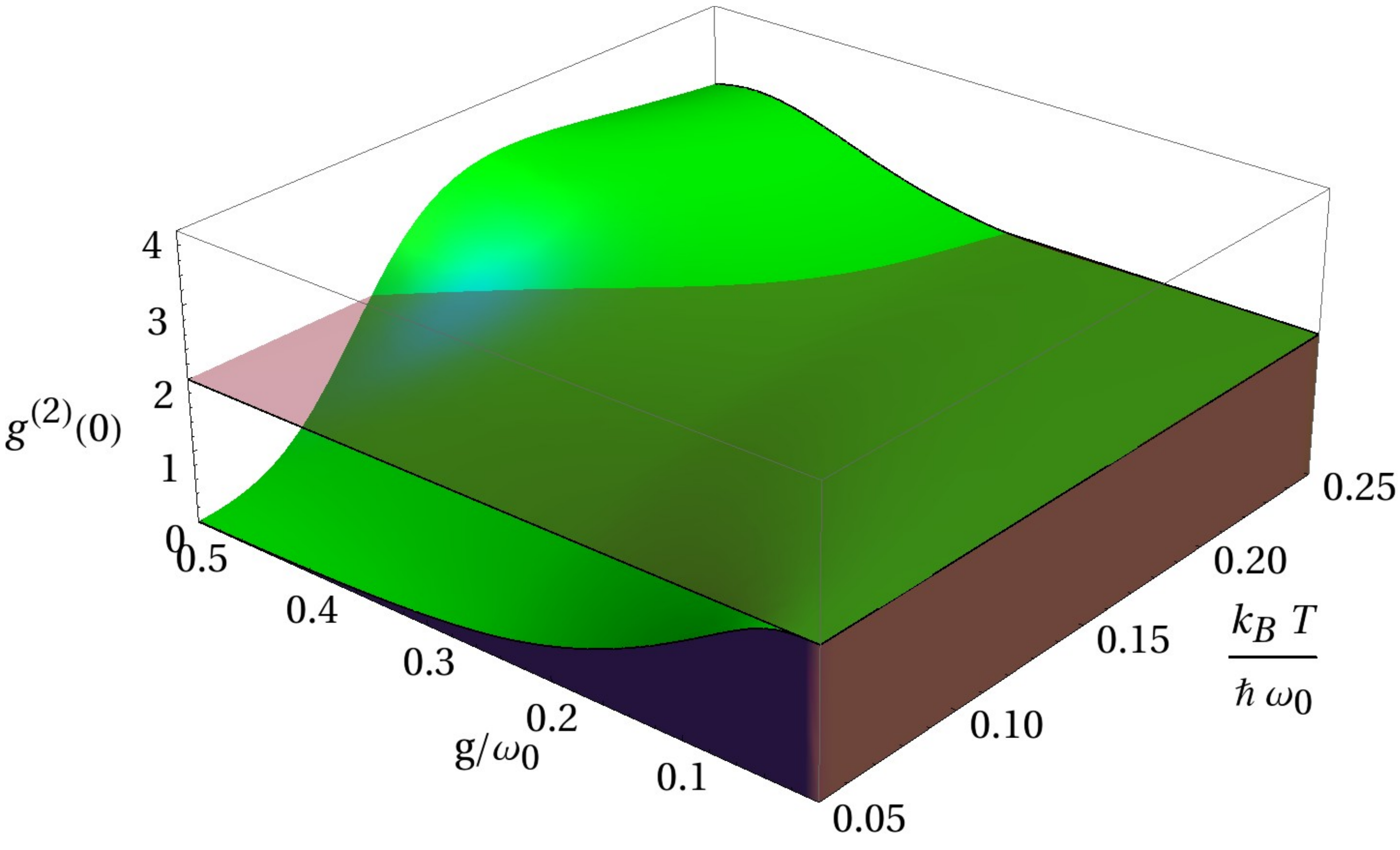}
\caption{(color online) $g^{(2)}(0)$ calculated for two TLS coupled with a single cavity mode.}
\label{fig:2TLS}
\end{figure}
%========================================================================
%========================================================================
\begin{figure}
\centering
\includegraphics[height=53mm]{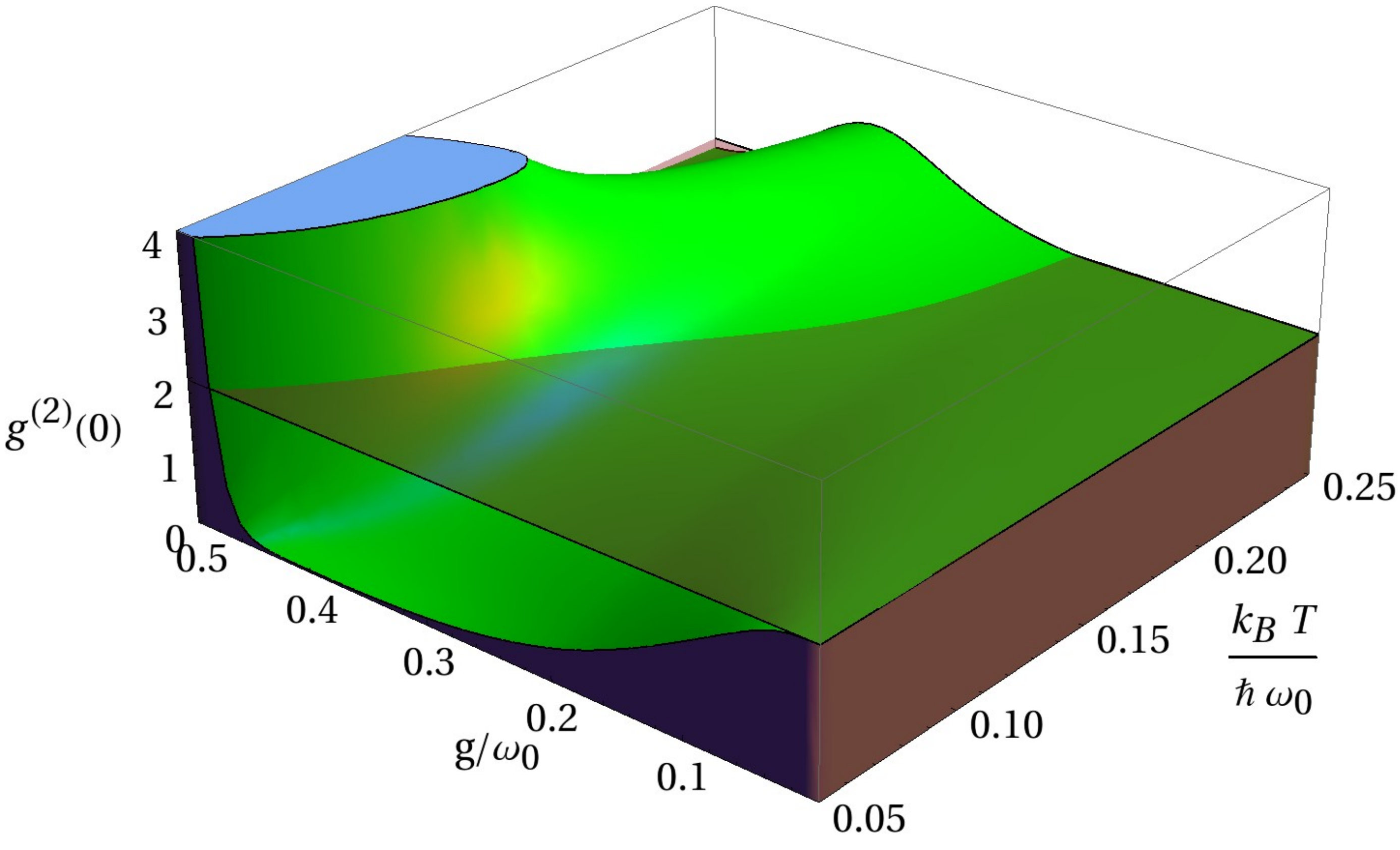}
\caption{(color online) $g^{(2)}(0)$ calculated for three TLS coupled to a single cavity mode.}
\label{fig:3TLS}
\end{figure}
%========================================================================
%========================================================================
\begin{figure}
\centering
\includegraphics[height=53mm]{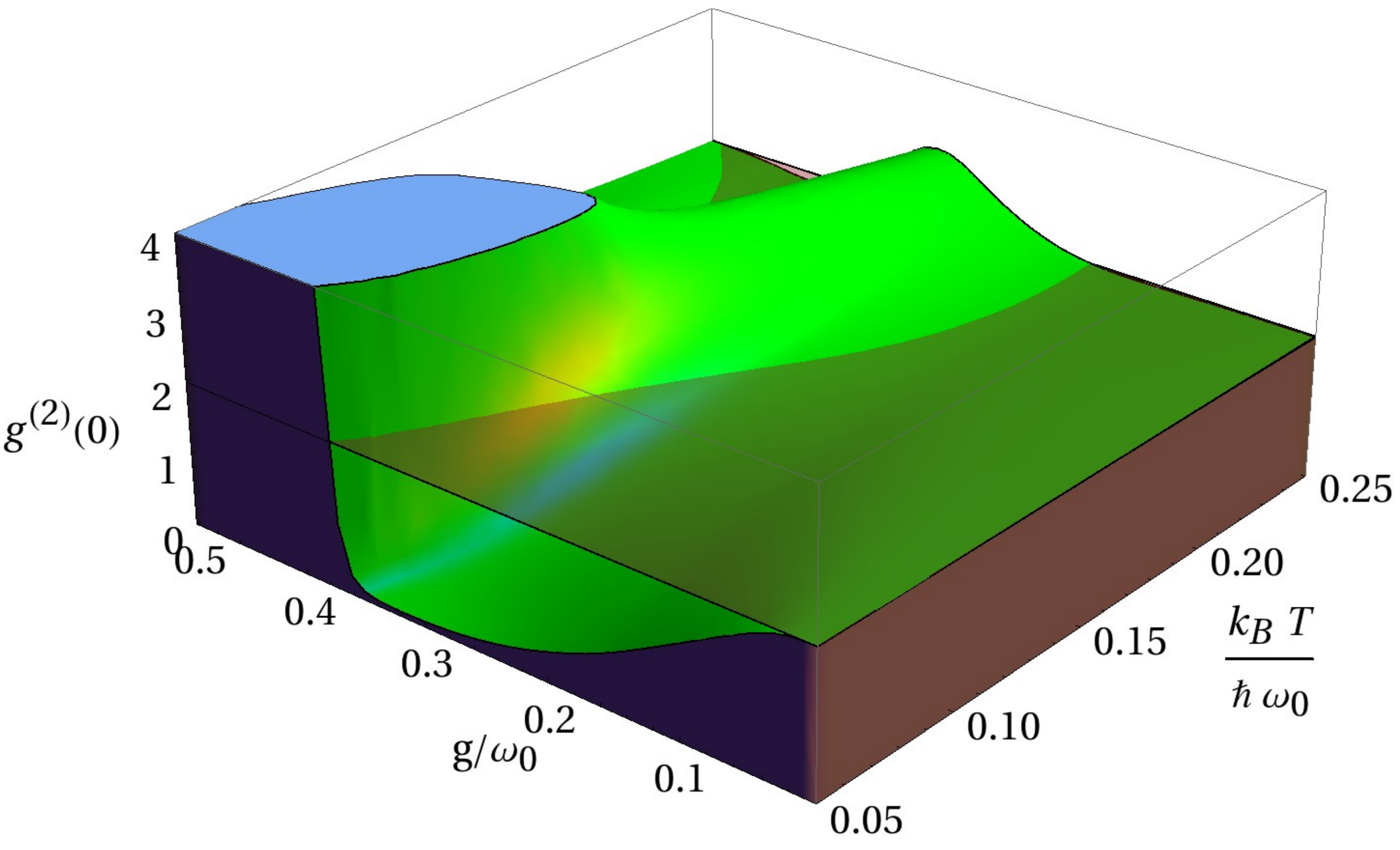}
\caption{(color online) $g^{(2)}(0)$ calculated for four TLS coupled to a single cavity mode.}
\label{fig:4TLS}
\end{figure}
%========================================================================
Figs. \ref{fig:2TLS}, \ref{fig:3TLS} and \ref{fig:4TLS} show the feasibility to reach a robust quantum regime, characterized by the strong antibunching, that persists even in presence of many TLSs. It is worth to note the onset of a significant superbunching for higher coupling strenghts. This is due to the degeneracy of the involved energy levels, that increases with the number of TLSs.
%========================================================================
\begin{figure}
\centering
\includegraphics[height=53mm]{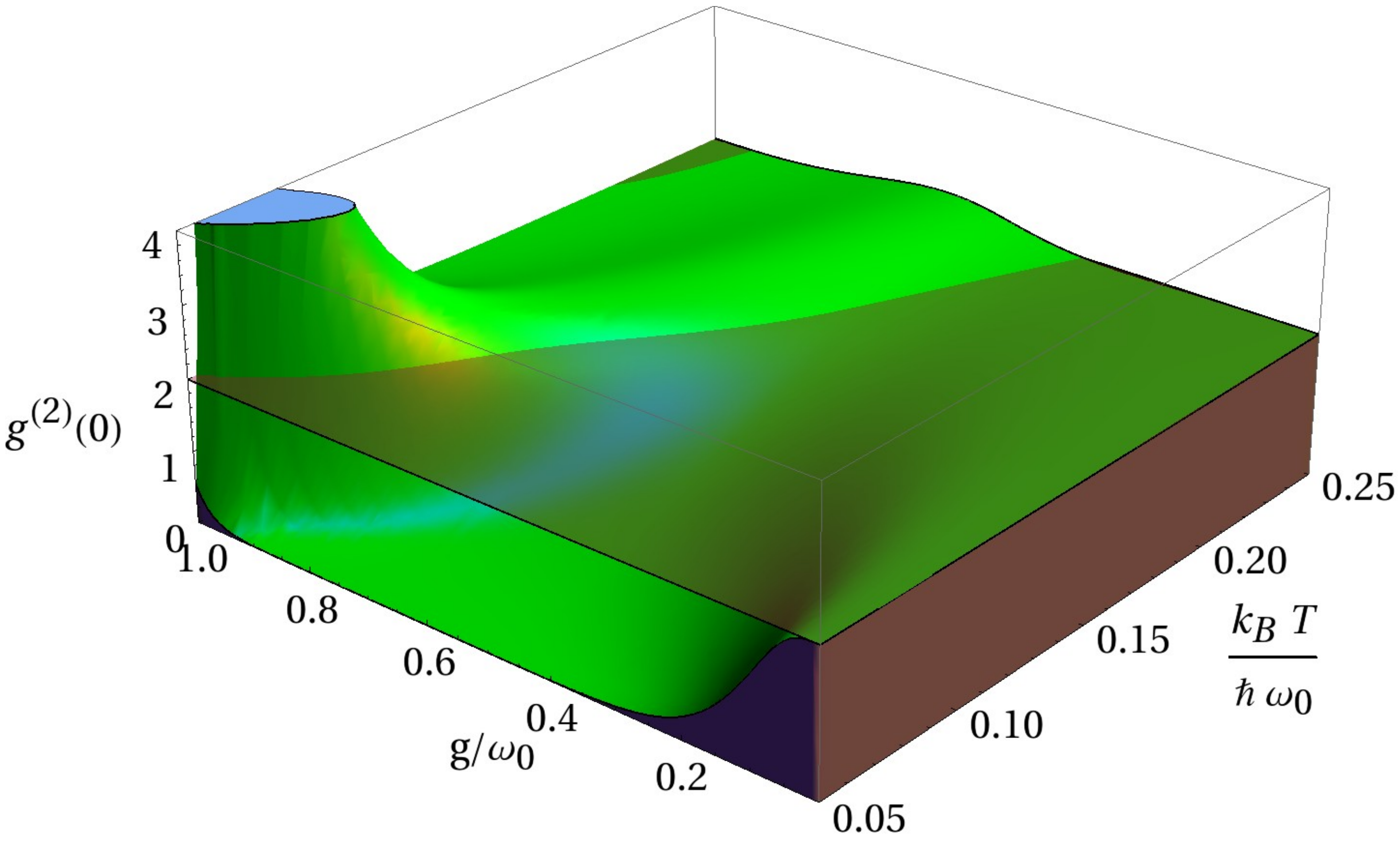}
\caption{(color online) $g^{(2)}(0)$ calculated for a TLS coupled to two cavity modes.}
\label{fig:2cav1TLS}
\end{figure}
%========================================================================

Fig. \ref{fig:2cav1TLS} in turn shows the thermal $g^{(2)}(0)$ calculated for a single TLS coupled with two cavity modes as described by the Hamiltonian
$H = \omega_{\rm 0}^{(1)} a^{\dagger}_{1} a_{1} + \omega_{\rm 0}^{(2)} a^{\dagger}_{2} a_{2} + \omega_{\rm x} \sigma^{+} \sigma^{-} + [g^{(1)} ( a_{1} + a^{\dagger}_{1}) +g^{(2)} ( a_{2} + a^{\dagger}_{2})] (\sigma^{-} + \sigma^{+})$.
In particular, for this latter calculation we used a second bare cavity mode with frequency $\omega_{\rm 0}^{(2)} = 2 \omega_{\rm 0}^{(1)}$ and a coupling with strength $g^{(2)} = 2 g^{(1)}$.
Such ratios of coupling strength appear for example in circuit QED experiments \cite{Niemczyk}. There is an enhancement of the antibunching even though we consider a second bosonic mode. The explanation of such a scenario is that the presence of the second cavity mode is compensated by the stronger coupling that effectively increase the nonlinarity of the whole system.

\paragraph{Acknowledgements -}
This work is part of the Emmy Noether project HA 5593/1-1 and the CRC 631, both funded by the German Research Foundation, DFG.

%%%%%%%%%%%%%%%%%%%%%%% References %%%%%%%%%%%%%%%%%%%%%%%%%


\begin{thebibliography}{99}
\bibitem{note} The expectation value of a generic one-time operator ${\cal O}(t)$ can be calculated as $\text{Tr}[{\cal O} \rho(t)]$.
\bibitem{expl} Note that, motivated by our photodetection theory, we always number the dressed states according to increasing energies.
\bibitem{step}
Our calculations correspond to the regime where the delay $\tau$ is large compared to the detector response time. Longer detector response times would
smoothen the jump of $g^{(2)}_{\rm 21,10}(\tau)$ at $\tau =0$, c.f. \cite{GonzalezTudela}.
\end{thebibliography}
\end{document}